\documentclass[preprintnumbers,
amsmath,amssymb,floatfix,10pt,prd,onecolumn,
superscriptaddress,longbibliography,nofootinbib]{revtex4-2}

\usepackage{graphicx}
\usepackage{dcolumn}
\usepackage{bm}
\usepackage[utf8]{inputenc}
\usepackage{textgreek}
\usepackage{palatino}
\usepackage{subcaption}
\usepackage{ragged2e}
\usepackage{xcolor}
\usepackage{hyperref}
\hypersetup{           
	colorlinks=true,                
	breaklinks=true,                
	urlcolor= blue,                
	linkcolor= magenta,                
	bookmarksopen=false,
	filecolor=black,
	citecolor=red,
	linkbordercolor=blue}
\usepackage{orcidlink}

\begin{document}


\title{Cosmological Realization of Baryon Asymmetry in $ f(R, G_{\mu\nu}T^{\mu\nu}) $ Gravity}


\author{Kalyan Malakar\orcidlink{0009-0002-5134-1553}}
\affiliation{Department of Physics, Dibrugarh University, Dibrugarh, 786004, Assam, India}
 \affiliation{Department of Physics, Silapathar College, Dhemaji, 787059, Assam, India}
 \email{kalyanmalakar349@gmail.com}
 
 \author{Rajdeep Mazumdar\orcidlink{0009-0003-7732-875X}}
\affiliation{Department of Physics, Dibrugarh University, Dibrugarh, 786004, Assam, India}
 \email{rajdeepmazumdar377@gmail.com}
 
\author{Kalyan Bhuyan\orcidlink{0000-0002-8896-7691}}
\affiliation{Department of Physics, Dibrugarh University, Dibrugarh, 786004, Assam, India}
\affiliation{Theoretical Physics Division, Centre for Atmospheric Studies, Dibrugarh University, Dibrugarh, 786004, Assam, India}
 \email{kalyanbhuyan@dibru.ac.in}


\begin{abstract}
This work investigates the mechanism of gravitational baryogenesis (GB) under the formalism of $f(R, G_{\mu\nu}T^{\mu\nu})$ gravity, where $R$ denotes the Ricci scalar, $G_{\mu\nu}$ is the Einstein tensor and $T^{\mu\nu}$ represents the energy--momentum tensor. $f(R, G_{\mu\nu}T^{\mu\nu})$ model is considered to evaluate the baryon-to-entropy ratio (BnER), which is subsequently compared against the observational limits. The results obtained exhibit compatibility with the estimated matter imbalance. Moreover, the analysis is extended to generalized GB case, resulting in outcomes that closely match empirical bounds. The findings reveal that the $f(R, G_{\mu\nu}T^{\mu\nu})$ formulation yields a viable theoretical setting for explaining the detected matter-antimatter disparity of the universe, highlighting its relevance in early cosmic evolution. To further validate the models, a chi-square ($\chi^2$) analysis of the Hubble parameter, $H(z)$, and distance modulus, $\mu(z)$, is performed, confirming their consistency with current cosmological observations. A comparative assessment simultaneously with the $\Lambda$CDM paradigm demonstrates a satisfactory level of agreement between the proposed model and cosmological observations from CC and Pantheon+SH0ES datasets.
\end{abstract}


\maketitle


\section{Introduction}
\label{sec1}

Among the earliest theoretical insights into the nature of matter, Paul Dirac proposed that the universe should contain anti-baryons, or antimatter, as a natural counterpart to ordinary matter \cite{Anderson1933Mar,Dirac1928Feb}. This prediction laid the groundwork for one of the most profound and longstanding unresolved issues in contemporary cosmology concerning the measured matter-antimatter disparity in the observable universe \cite{cohen1998matter}. The theoretical framework developed to explain this imbalance is known as baryogenesis \cite{dine2003origin}, a set of physical processes believed to have unfolded in the very early universe \cite{sakharov1998violation}, generating a net excess of baryonic matter over its antimatter counterpart \cite{bennett2003microwave,dolgov2001matter,dolgov1988cosmology}.

Of the entire cosmic content of the cosmos, dark matter accounts for approximately 31.7 \%, manifesting in two distinct forms - baryonic and non-baryonic \cite{dine2003origin}. Baryonic dark matter constitutes roughly 4.9 \% of the universe and encompasses all visible, luminous matter \cite{weinberg1995quantum,bennett2003microwave,Peebles2003Apr}. Non-baryonic dark matter, by contrast, neither emits nor interacts with em radiation, yet its presence is inferred through its gravitational influence \cite{Kolb2018Mar}.

From the standpoint of modern cosmology, a symmetric origin of the universe would be expected to produce equal quantities of matter and antimatter, yielding no net baryonic excess \cite{dine2003origin,weinberg1995quantum}. Yet, this expectation stands in sharp contrast to every piece of empirical evidence available, from everyday observations to the most precise cosmological measurements \cite{bennett2003microwave,Aghanim2020Sep}. The universe we inhabit is overwhelmingly composed of matter, with virtually no trace of large-scale antimatter domains \cite{steigman1976observational,dolgov2001matter,burles2001big}.

This conclusion is robustly supported by multiple observational evidences. The precise results of the Cosmic Microwave Background (CMB) \cite{bennett2003microwave}, the success of BBN in predicting the primordial abundances of light elements \cite{burles2001big}, and the conspicuous absence of high-energy $\gamma$-rays stemming from baryon-antibaryon annihilation all converge on the same conclusion: the universe harbors a profound asymmetry between matter and antimatter \cite{Zeldovich1968Mar}. Addressing this fundamental tension is the central objective of baryogenesis theories which seek to identify the physical mechanisms responsible for breaking the initial matter-antimatter symmetry.

A quantitative characterization of this asymmetry is conventionally expressed through a dimensionless parameter that captures the excess of baryons over antibaryons relative to the photon number density \cite{dine2003origin,davoudiasl2004gravitational}:

\begin{equation}
\frac{\eta_B}{\gamma} = \frac{n_B - n_{\bar{B}}}{\gamma}
\label{eq:baryon_to_photon_ratio}
\end{equation}

Andrei Sakharov outlined three essential criteria that any viable mechanism fulfill to dynamically produce a net baryon asymmetry from an initially symmetric state \cite{sakharov1998violation}. The initial criterion necessitates the presence of physical processes that violate baryon number conservation, the next criterion demands the non-conservation of both $\mathcal{C-}$ \& $\mathcal{CP-}$ invariance, and the third condition stipulates that these processes must occur out of thermal equilibrium.

\textbf{Baryon number violation:}
Let $J_B^\mu$ denote the baryon current. If baryon number is conserved, it satisfies the continuity equation \cite{Cheng1984}:

\begin{equation}
\partial_\mu J_B^\mu = 0 .
\end{equation}

The total baryon number of the system is defined as
\begin{equation}
B = \int d^3x\, J_B^0 .
\end{equation}

Consequently, assuming negligible surface terms:

\begin{equation}
\frac{dB}{dt} = \int d^3x\, \partial_0 J_B^0 .
\end{equation}

A necessary condition for inducing baryon excess is therefore the violation of baryon number conservation,
\begin{equation}
\partial_\mu J_B^\mu \neq 0 ,
\end{equation}
which allows processes with $\Delta B \neq 0$.

\textbf{C and CP violation:}
Consider a generic transition in which a particle species $A$ decays into a final state $f$ carrying baryon number $B_f$. Let $\Gamma(A\rightarrow f)$ denote the decay rate for this process and $\Gamma(\bar{A}\rightarrow \bar{f})$ the decay rate for the corresponding antiparticle decay. The $\mathcal{CP-}$ breaking coefficient can be defined as \cite{Riotto1999Dec}:

\begin{equation}
\epsilon =
\frac{\Gamma(A\rightarrow f) - \Gamma(\bar{A}\rightarrow \bar{f})}{\Gamma(A\rightarrow f) + \Gamma(\bar{A}\rightarrow \bar{f})}
\end{equation}

If charge conjugation and parity were exact symmetries, one would obtain $\epsilon=0$. A non-vanishing value of $\epsilon$ therefore signals the non-conservation under $\mathcal{C-}$ \& $\mathcal{CP-}$ transformations and allows the generation of unequal numbers of particles and antiparticles.

\textbf{Out-of-equilibrium thermal state:}
Finally, baryon-number generating processes must occur out of thermal equilibrium. In equilibrium the principle of detailed balance requires that the transition rates satisfy \cite{Riotto1999Dec}:

\begin{equation}
W_{i\rightarrow f}\, n_i^{eq} =
W_{f\rightarrow i}\, n_f^{eq},
\end{equation}

where $W_{i\rightarrow f}$ represents the probability of transition per unit time whereas $n^{eq}$ denotes the equilibrium number density. Under this condition any baryon number produced by a reaction is compensated by the inverse process, leading to:

\begin{equation}
\left(\frac{dn_B}{dt}\right)_{eq}=0 .
\end{equation}

A rich landscape of theoretical proposals has been put forward over the decades in an effort to describe the inferred baryon asymmetry. Some of the most widely studied frameworks are Affleck–Dine baryogenesis \cite{stewart1996affleck,dine2003origin}, which exploits the scalar-field-driven evolution that carries the baryon number in the early cosmos; electroweak baryogenesis \cite{trodden1999electroweak}, which attributes the asymmetry to non-equilibrium processes occurring at the electroweak phase transition; and Grand Unified Theories (GUTs) \cite{kolb1996grand}. Further proposals include spontaneous baryogenesis \cite{brandenberger2003spontaneous,takahashi2004spontaneous}, wherein a dynamically evolving scalar field spontaneously breaks $\mathcal{CPT}$ invariance to favor baryon production; mechanisms based on leptogenesis \cite{Luty1992Jan,Barbieri2000May}; and more exotic scenarios such as baryogenesis through the evaporation of primordial black holes \cite{ambrosone2022towards}. Each of these frameworks approaches the problem from a distinct theoretical perspective, yet they share the common goal of identifying a consistent and physically motivated origin for the matter-dominated universe we observe today.

The gravitational baryogenesis mechanism was devised by Davoudiasl et al. \cite{davoudiasl2004gravitational}, in which the required matter–antimatter asymmetry can emerge naturally through the coupling of $\partial_{\mu} R$ to baryonic matter. The central ingredient of this framework is a non-conserving $\mathcal{CP}$ dynamics which allows a preferential production of baryons over antibaryons \cite{davoudiasl2004gravitational}. This interaction is given explicitly by:

\begin{equation}
\frac{1}{M_*^2}
\int (\partial_{\mu}R) J^{\mu}\ \sqrt{-g} d^4x \,\;,
\label{eq:coupling_equation}
\end{equation}
where, \(M_*\) stands for the cut-off scale of the EFT.

Originally, the gravitational baryogenesis mechanism encountered a fundamental limitation: in a cosmos dominated by radiations, the time derivative of $ R \ (\dot{R}) $ vanishes identically, rendering the mechanism incapable of generating any net baryon asymmetry in this epoch \cite{lambiase2006baryogenesis}. This shortcoming has motivated a number of authors to seek extended or modified versions of the original framework. Davoudiasl et al. $(2004)$ addressed this issue by demonstrating that the interplay between $ \dot{R} $ and $J_B^{\mu}$ spontaneously violates the $\mathcal{CPT-}$ symmetry which can successfully amplify the baryonic surplus to that of the observed bounds \cite{davoudiasl2004gravitational}. Subsequent work by Oikonomou $(2016)$ \cite{Oikonomou2016Mar}, Oikonomou and Saridakis $(2016)$ \cite{Oikonomou2016Dec}, Odintsov and Oikonomou $(2016)$ \cite{Odintsov2016Sep}, Nozari et. al. $(2018)$ \cite{nozari2018baryogenesis}, Atazadeh $(2018)$ \cite{Atazadeh2018Jun} and Baffou et al. $(2019)$ \cite{baffou2019f} explored a complementary strategy, showing that the Friedmann equations governing cosmological evolution are subject to non-trivial modifications when the classical geometric description is replaced by a modified gravity theory. In these settings, the modified Friedmann dynamics introduce corrections that reactivate the gravitational baryogenesis mechanism, enabling a non-zero baryon asymmetry to be generated even in an era of radiation background, a result that cannot be achieved within the standard GR framework.

This observation has brought considerable attention to alternative geometric formulations of gravity, which modify the underlying description of spacetime while reproducing the observational successes of GR. Among the most prominent of these are TEGR \cite{SwagatMishra2026Jan}, which reformulates gravitational dynamics entirely in terms of torsion scalar $(T)$ and STEGR \cite{Mishra2024Apr}, which employs the idea of non-metricity, captured by the scalar $Q$. The theoretical landscape has been further broadened by a variety of hybrid and extended constructions, including $f(R, L_m)$ gravity\cite{jaybhaye2023baryogenesis}, $f(T, \phi)$ \cite{Sultan2025May}, $f(T, L_m)$\cite{Cruz2026Feb}, $f(Q, L_m)$ \cite{Samaddar2025Mar}, $f(R,L_m,T)$ \cite{Malakar2026Jun}, and Einstein-aether gravity \cite{Sultan2025Mar} each offering distinct extensions of the geometric framework underlying gravitational physics.

The present work explores a novel theoretical extension of gravitational baryogenesis rooted in a geometry–matter coupling that is non-minimal in nature. Specifically, we examine the implications of introducing an interaction between $G_{\mu\nu}$ and $T_{\mu\nu}$ \cite{Marciu2023Oct}, a pairing that has not been investigated in the baryogenesis context. Building upon this coupling, we construct an extended gravitational model in which a general function encoding the matter–geometry interplay is incorporated directly into the gravitational action. The resulting framework thus naturally decomposes into two distinct contributions: a purely geometric sector governed by the Ricci scalar, and a second sector characterized by a new curvature invariant constructed from the contraction coupling. We aim to investigate whether this modified matter–geometry coupling can account for the observed gravitational baryogenesis conditions, thereby offering new insight into the emergence of the measured baryonic excess of the cosmos.

The manuscript is structured as follows. Section \ref{sec2} provides the foundational formulation of the \(f(R,\xi)\) gravity framework and derives the associated field equations. This section further outlines the essential aspects of the GB scheme. In Sections \ref{sec3} and \ref{sec4}, the GB formalism and its generalized extension within modified \(f(R,\xi)\) model is rigorously developed through a detailed analytical treatment. These sections further present graphical analyzes, together with an in-depth discussion of the resulting physical implications and major outcomes. In Section \ref{sec5}, the consistency of the parameter space obtained  with available cosmological observations is investigated under the constraints imposed by gravitational baryogenesis. The final Section \ref{sec6} summarizes the key results of the study and presents the concluding observations.

\section{\(f(R, G_{\mu\nu} T^{\mu\nu})\) gravity \& Field Equations}
\label{sec2}
\subsection{\(f(R, G_{\mu\nu} T^{\mu\nu})\) Gravity}

In this work, we explore the possible interaction between $G_{\mu\nu}$ and $T^{\mu\nu}$ as a means to probe modifications of standard gravitational dynamics. By incorporating such a coupling, we extend the underlying theoretical framework to examine gravitational baryogenesis. This is achieved by introducing a general functional term into the Lagrangian that captures the combined influence of matter and geometry. Consequently, the action for the gravity model with non-minimal matter–geometry coupling is given by \cite{Marciu2023Oct}:

\begin{equation}
S=
\int \sqrt{-g} \ d^4x \,[\ f(R) + h(\xi) ] + S_m
\label{eq:action_equation}
\end{equation}

In addition to the standard $f(R)$ contribution, we introduce a general function, $h(\xi)$, which encapsulates the non-minimal interaction between matter and geometry. This function is constructed from a scalar quantity ξ, defined through the contraction of the Einstein tensor with the energy–momentum tensor, such that:
\begin{equation}
\xi=G_{\mu\nu} T^{\mu\nu}
\label{eq:xi_term}
\end{equation}

To derive the governing dynamical equations associated with the framework of $ f(R, G_{\mu\nu} T^{\mu\nu}) $, the principle of least action is applied. Carrying out this variational procedure yields a class of generalized, Einstein-like field equations that govern the spacetime dynamics within the present framework and is given as:

\begin{equation}
T_{\mu\nu}^{fR} + T_{\mu\nu}^{h \xi} + T_{\mu\nu}^{M} = 0.
\label{eq:Variation_of_action}
\end{equation}

The tensors \(T_{\mu\nu}^{fR}\), \(T_{\mu\nu}^{h \xi}\) and \(T_{\mu\nu}^{M}\) denote the contributions to $T_{\mu\nu}$ tensor arising from the $ f(R) $ sector, the $ h(\xi) $ coupling term, and the matter component, respectively.

That consequently leads to a modified conservation equation as follows:

\begin{equation}
\nabla_{\mu} \left(T_{\mu\nu}^{fR} + T_{\mu\nu}^{h \xi} + T_{\mu\nu}^{M} \right) = 0,
\label{eq:conservation_relation}
\end{equation}

which is alternatively stated as continuity relation.

The quantity \(T_{\mu\nu}\) appearing in Eq.~\eqref{eq:xi_term} corresponds to the energy–momentum tensor, which is expressed as:  

\begin{equation}
T_{\mu\nu}=-2\frac{1}{\sqrt{g}}\frac{\delta (\sqrt{-g}L_{m})}{\delta g^{\mu\nu}} = -2\frac{\partial L_m}{\partial g^{\mu\nu}} + g_{\mu\nu} L_m.
\label{eq:energy-momentum_tensor}
\end{equation}

In this framework, $ L_m $ represents the Lagrangian density associated with the matter component, which encapsulates the physical content of all matter fields present in the $ f(R, G_{\mu\nu} T^{\mu\nu} ) $ theory. The trace of the $T_{\mu\nu}$ tensor is obtained through contraction of the metric tensor  with the contravariant tensor $T^{\mu\nu}$, and is defined as: $ T = g_{\mu\nu} T^{\mu\nu} $.

Performing variation of the action relative to the inverse metric leads to the Einstien-like field equation that generates the $T_{\mu\nu}$ tensors linked to the $ f(R) $ and $ h(\xi) $ sectors respectively, expressed as follows

\begin{equation}
T_{\mu\nu}^{fR} = g_{\mu\nu} f(R) + 2 \ \nabla_{\mu} \nabla_{\nu} f_R - 2 \ g_{\mu\nu} \Box f_R - 2 \ R_{\mu\nu} f_R \ ,
\label{eq:fR_energy-momentum_tensor}
\end{equation}

where, $f_R$ represents curvature scalar derivative of $f(R)$.

\begin{equation}
\begin{split}
T_{\mu\nu}^{h \xi} &= g_{\mu\nu}\, h(\xi) + h_{\xi} T R_{\mu\nu} - 2 h_{\xi} G_{\nu}^{\beta} T_{\mu\beta} 
- 2 h_{\xi} G^{\alpha}_{\mu} T_{\nu\alpha} - h_{\xi} R T_{\mu\nu} - \square \left( h_{\xi} T_{\mu\nu} \right) + \nabla_{\alpha} \nabla_{\mu} \left( h_{\xi} T_{\nu}^{\alpha} \right) + \nabla_{\alpha} \nabla_{\nu} \left( h_{\xi} T_{\mu}^{\alpha} \right) \\ 
 & - g_{\mu\nu} \nabla_{\alpha} \nabla_{\beta} \left( h_{\xi} T^{\alpha\beta} \right) + g_{\mu\nu} \square \left( h_{\xi} T \right) - \nabla_{\mu} \nabla_{\nu} \left( h_{\xi} T \right) - 2 h_{\xi} \Sigma_{\mu\nu} \ ,
\end{split}
\label{eq:h_xi_energy-momentum_tensor}
\end{equation}

where, $h_{\xi}$ stands for the contraction scalar $(\xi)$ derivative of $h(\xi)$. 

where, we define the contracted quantity \(\Sigma_{\mu\nu}\) as follows:

\begin{equation}
\begin{split}
\Sigma_{\mu\nu} &= G^{\alpha\beta} \frac{\delta T_{\alpha\beta}}{\delta g^{\mu\nu}} \\ 
& = \frac{1}{2} G^{\alpha\beta} g_{\alpha\beta} \left(g_{\mu\nu} L_m - T_{\mu\nu} \right) - G_{\mu\nu} L_m - 2 G^{\alpha\beta} \frac{\delta^2 L_m}{\delta g^{\mu\nu} \delta g^{\alpha\beta}} \ ,
\end{split}
\label{eq:Sigma-mu_nu}
\end{equation}

where, the quantity $\frac{\delta T_{\alpha\beta}}{\delta g^{\mu\nu}}$ is derived as follows:

\begin{equation}
\frac{\delta T_{\alpha\beta}}{\delta g^{\mu\nu}}=\frac{1}{2} g^{\alpha\beta} L_m \ g_{\mu\nu} + \frac{\delta g_{\alpha\beta}}{\delta g^{\mu\nu}} L_m - 2 \frac{\delta^2 L_m}{\delta g^{\mu\nu} \delta{g^{\alpha\beta}}} - \frac{1}{2} g_{\alpha\beta} T_{\mu\nu}.
\label{eq:trace_derivative-of_metric}
\end{equation}

The unusual appearance of the conservation relation Eq. \eqref{eq:energy-momentum_tensor} is due to the inclusion of contraction term $h(\xi)$ in the gravity theory $f(R, G_{\mu\nu}T^{\mu\nu})$. It can be easily shown that the Eq. \eqref{eq:energy-momentum_tensor} reduces to the familiar form conservation condition (\(\nabla_{\mu}T^{\mu\nu}=0\)) when consider the limiting cases, \(f(R)=R\) and \(f(\xi)=0\).

To characterize the cosmic-scale spatial and directional uniformity of the cosmos, we adopt the FLRW metric ansatz, written as \cite{rasanen2015new}:

\begin{equation}
ds^2 = -dt^2 + a^2(t) \ \delta_{\gamma\delta} \ dx^{\gamma} dx^{\delta} .
\label{eq:flat_FLRW_metric}
\end{equation}

The choice of FLRW metric is well-supported by observational evidence from large-scale structure (LSS), CMB, and baryon acoustic oscillation (BAO) surveys \cite{l2017model,jimenez2019measuring,foidl2024lambda}, which together enhance our perceptions of spacetime expansion history and the current epoch of accelerated expansion. 

\subsection{Field Equations}
Within the cosmological setting, matter is commonly treated under the ideal fluid description, wherein the fluid is assumed to be isotropic. Under this assumption, the spatial distribution of energy density and momentum propagation across the spacetime manifold is fully captured by the \(T_{\mu\nu}\) tensor, which takes the following standard form \cite{Wald1984Jan}:

\begin{equation}
T_{\mu\nu}=pg_{\mu\nu} + (p+\rho)u_{\mu}u_{\nu},
\label{eq:T_as_perfect_fluid}
\end{equation}

In the expression \eqref{eq:T_as_perfect_fluid},  \(\rho\) represents the energy density, \(p\) is the isotropic pressure, \(u^{\mu}\) and  is the four-velocity vector of the fluid, which satisfies the normalization relation \(u^{\mu} u_{\mu}=1\), consistent with the signature of the spacetime metric.

Employing Eq. \eqref{eq:Variation_of_action}, the generalized Friedmann equations may be derived for the metric presented in Eq. \eqref{eq:flat_FLRW_metric}, yielding the following result: 

\begin{eqnarray}
&& f(R) -6 f_{R} (\dot{H} + H^2) + 6 H f_R = \rho - h(\xi) - 6 \rho h_{\xi} \dot{H},
\label{eq:friedmann_equation1} \\&&
f(R) -2 f_R (\dot{H} +3 H^2) + \ddot{f_R} + 4 H \dot{f_R} = -p_{\xi}.
\label{eq:friedmann_equation2}
\end{eqnarray}

where, 

\begin{align*}
    p_{\xi} = h(\xi) -2 h_{\xi} (\rho (\dot{H} 3 H^2) + \dot{\rho} H) - 6 H^2 \rho (2 \rho \dot{H} + \dot{\rho} H) h_{\xi\xi}
\end{align*}

A single overdot and a double overdot placed above a given quantity denote its first-order and second-order differentiation in relation to cosmic time, respectively.

In a spatially flat FLRW background, the quantities $R$ and $\xi$ can be expressed in a specific algebraic form, which is given by the following relations. These expressions are derived directly from the geometry of the flat FLRW metric and will be used in the subsequent analysis.

\begin{equation}
R=6(2H^2+\dot{H}).
\label{eq:ricci_scalar}
\end{equation}

and,

\begin{equation}
\xi = 3 H^2 \rho.
\label{eq:coupling_scalar}
\end{equation}

In this context, the equation‑of‑state (EoS) parameter, denoted by $(\omega)$, is introduced as the ratio between the pressure $(p)$ and the energy density $(\rho)$. This dimensionless quantity characterizes how the pressure of the cosmic fluid relates to its energy content and plays a central role in determining the dynamical evolution of the cosmos.

\subsection{Insights into the Theory of Gravitational Baryogenesis}

The prevalence of matter in the visible universe constitutes a fundamental challenge the  cosmological results Standard Model. Theoretical predictions initially suggest a primordial state of baryonic symmetry; however, empirical evidence from BBN and CMB anisotropies \cite{burles2001big, bennett2003microwave} definitively points toward a non-vanishing net baryon number. Furthermore, the lack of detectable gamma-ray signals from large-scale annihilation events \cite{cohen1998matter} confirms that the cosmos is overwhelmingly composed of matter. This discrepancy is formally represented by the BnER ($\eta_B/s$), which acts as a dimensionless proxy for the observed asymmetry \cite{lambiase2006baryogenesis}:

\begin{equation} 
\frac{\eta_B}{s} = \frac{n_B - n_{\bar{B}}}{s}. 
\label{eq:baryon-asymmetry-def} 
\end{equation}

In this expression, $n_B$ and $n_{\bar{B}}$ correspond to the respective particle densities of baryonic matter and antimatter, whereas, $s$ represents the entropy of the universe. Constraints from BBN \cite{burles2001big,White2022Apr} yield a BnER of $\frac{n_B}{s} = (7.3 \pm 2.5)\times 10^{-11}$, while analyses of the CMB anisotropies \cite{White2022Apr} indicate $\frac{n_B}{s} = (6.19 \pm 0.14)\times 10^{-10}$ at the confidence level of 95\%. WMAP and Planck missions \cite{White2022Apr,bennett2003microwave,Aghanim2020Sep} suggest the asymmetry to be: $(9.2 \pm 1.1) \times 10^{-11}$ and $(8.59 \pm 0.11) \times 10^{-11}$ respectively. For our framework, the numerical magnitude of baryon asymmetry is taken as $\frac{n_B}{s} = 9.2^{+0.6}_{-0.4}\times 10^{-11}$, to constrain the free model parameters.

Motivated by the $\mathcal{CP-}$ violating interaction term proposed by Davoudiasl et al.~\cite{davoudiasl2004gravitational} in Eq. \eqref{eq:coupling_equation}, a similar coupling term can be devised for $f(R, \xi)$ gravity that violates $\mathcal{CP}$ conservation, which takes the following form:

\begin{equation}
\frac{1}{M_*^2} \int d^4x \, \sqrt{-g}\,  \ \partial_\mu (R+\xi) \ J^\mu.
\label{eq:modified_interaction_term_GB}
\end{equation}

As the Universe expands, its temperature $(T)$ decreases, and once it falls below a critical value called, decoupling Temperature, $(T_D)$, baryon–number–violating interactions effectively switch off. The baryon asymmetry generated until $(T_D)$ then freezes, resulting in a constant BnER as a relic of the early cosmos. The framework of GB scripts this imbalance through an effective interaction between the Ricci curvature derivative and the baryon current, as in Eq. \eqref{eq:coupling_equation}, which induces a $\mathcal{CPT-}$ violating chemical potential in thermal equilibrium. In this setup, the resulting BnER is given by \cite{lambiase2006baryogenesis,Malakar2026Jun}:

\begin{equation}
\frac{\eta_B}{s} \simeq - \frac{15 g_b}{4 \pi^2 g_*} \frac{\dot{R}}{M_*^2 T_D},
\label{eq:ratio_for_R}
\end{equation}

here, $g_b$ counts the intrinsic baryonic degrees of freedom, $g_*= \frac{45 s}{2\pi^2 (T_D)^3}$ represents the effective  d.o.f of the relativistic species that constitutes the entropy of the universe, and $\dot{R}$ is the time differentiation of R evaluated at the decoupling time, $t_D$.

In a similar manner, based on the coupling term in Eq. \eqref{eq:modified_interaction_term_GB}, the resulting BnER, under the $f(R,\xi)$ framework, is given as:

\begin{equation}
\frac{\eta_B}{s} \simeq - \frac{15 g_b}{4 \pi^2 g_*} \frac{(\dot{R} + \dot{\xi})}{M_*^2 T_D},
\label{eq:ratio_for_R_and_xi}
\end{equation}

We set $\xi = \xi_0 \, \bar{\xi}$, where $\bar{\xi}$ and $R$ have units of $\text{GeV}^6$ and $\text{GeV}^2$, respectively. The parameter $\xi_0$ is chosen to carry the unit $\text{GeV}^{-4}$, ensuring that $\xi$ itself has units of $\text{GeV}^2$. Consequently, in equation \eqref{eq:ratio_for_R_and_xi}, the sum $R + \xi$ also possesses units of $\text{GeV}^2$. Here, $\xi_0$ is a unit magnitude parameter introduced solely for dimensional consistency.

This process is responsible for producing the baryon imbalance seen in the Universe. The coupling term \eqref{eq:ratio_for_R_and_xi} is crucial in inducing $\mathcal{CP}$ and $\mathcal{CPT}$ symmetry breaking, ultimately favoring matter creation.

During the era governed by radiation, the temperature and energy density are related by a standard statistical connection. This relationship illustrates the dynamics of the energy density scales with temperature in the early cosmos, which is characterized by a series of quasi-equilibrium states, offering a deeper understanding of the cosmic transition during that period, as follows \cite{piattella2018lecture,Mukhanov2005Nov}:

\begin{equation}
\rho = \frac{\pi^2}{30} g_* T^4.
\label{eq:energy_density}
\end{equation}

As expansion progresses, subsequent cooling beyond $T_D$ triggers phase transitions that facilitate the violation of $\mathcal{CP}$ necessary for baryogenesis. This sequence of events is vital for the synthesis of fundamental particles and the eventual formation of LSS \cite{sugamoto1995baryon, balaji2005dynamical, huber2023baryogenesis}.

In this analysis, we model the cosmic expansion using a scale factor of power-law form \cite{malakar2025f}:

\begin{equation} 
a(t) = p_0 t^{\varphi}, 
\label{eq:power_law_a} 
\end{equation}

where the growth index $\varphi$ is a positive real quantity defined by the EoS parameter $(\omega)$ as $\varphi = \frac{2}{3(1+\omega)}$. $p_0$ stands for the magnitude of $a(t)$, at $t=0$. This power-law formulation provides a robust and mathematically tractable framework, allowing for the investigation of baryogenesis across various cosmological epochs by adjusting the value of $\varphi$. It should be noted that during the radiation-dominated phase of GR, the BnER becomes zero. Hence, our study examines whether the $f(R,\xi)$ gravity framework is capable of replicating the matter favoring scenarios of the early universe in that same era.

\section{Gravitational Baryogenesis in \(f(R, \xi)\)}
\label{sec3}

To obtain analytical solutions of Eqs. \eqref{eq:friedmann_equation1} and \eqref{eq:friedmann_equation2}, it is required to specify a concrete form of \(f(R,\xi)\) gravity rather than working with general formulation of the gravity theory.  We consider the specific form \(f(R, \xi)= R^2 + \beta \ \xi^{\alpha} \) , originally proposed in ref. \cite{Marciu2023Oct}. The purpose of this investigation is to examine whether this framework can successfully explain the observed baryon imbalance through a \(\mathcal{CP}\)-non-conserving interplay. The model parameters are confined to cosmologically admissible bounds to ensure that the theoretically predicted baryon-to-entropy ratio (BnER) remains consistent with results suggested  by observations. The analysis proceeds by first expressing  \(t_D\) as a function of  \(T_D\), and subsequently computing the BnER for the selected model using Eq. \eqref{eq:ratio_for_R_and_xi}.

\subsection{\(f(R, \xi)\) Model}

The analysis is carried out using the \(f(R, \xi)\) model given by \cite{Marciu2023Oct}:
\begin{equation}
f(R, \xi) = R^2 + \beta \ \xi^{\alpha} ,
\label{eq:f_form}
\end{equation}
here, \(\alpha\) and \(\beta\) represent the model parameters.

By combining Eqs. \eqref{eq:friedmann_equation1}, \eqref{eq:ricci_scalar}, \eqref{eq:power_law_a} and \eqref{eq:f_form}, the following expression for the energy density $(\rho)$ is obtained:

\begin{equation}
\rho = \left(\frac{3^{\alpha } (1-4 \alpha ) \beta }{(2 t)^{2 \alpha }}\right)^{\frac{1}{1-\alpha }}
\label{eq:energy_density_derived}
\end{equation}

A relation between \(T_D\) and \(t_D\) can be obtained by comparing Eqs. \eqref{eq:energy_density} and \eqref{eq:energy_density_derived}, yielding:

\begin{equation}
t_D = \frac{1}{2} \left(\frac{3^{\alpha } (1-4 \alpha ) \beta }{\left(\frac{1}{30} \pi ^2 g_* \ T_D^4\right)^{1-\alpha }}\right)^{\frac{1}{2 \alpha }}.
\label{eq:tD_for_GB}
\end{equation}

By inserting Eqs. \eqref{eq:ricci_scalar}, \eqref{eq:energy_density_derived}, and \eqref{eq:tD_for_GB} into Eq. \eqref{eq:ratio_for_R_and_xi}, the BnER can be written as:

\begin{equation}
\frac{\eta_B}{s} = -\frac{3^{2-\frac{3}{2 \alpha }} \left(r^{\alpha -1} \pi ^{-2 (1-\alpha )} (1-4 \alpha ) \beta \right)^{-\frac{3}{2 \alpha }} \left(-3^{\alpha } (4 \alpha -1) \beta  \left(3^{\frac{1}{2 \alpha }} \left(r^{\alpha -1} \pi ^{-2 (1-\alpha )} (1-4 \alpha ) \beta \right)^{\frac{1}{2 \alpha }}\right)^{-2 \alpha }\right)^{\frac{1}{1-\alpha }}}{r \ \pi ^2 (\alpha -1)},
\label{eq:ratio_expsn_GB}
\end{equation}

where,

$$r=16.96 \times 10^{37}$$

As discussed in \cite{davoudiasl2004gravitational}, the coupling factor of the type appearing in Eq. \eqref{eq:coupling_equation} emerges intrinsically in the low-energy EFT stemming from QFT in non-flat background, provided that \(M_*\) is of the order of the reduced Planck mass, \(M_*=\frac{m_{Pl}}{\sqrt{8\pi}}\), where, \((M_{Pl})\) stands for the Planck mass \cite{sahoo2020gravitational,nozari2018baryogenesis}. The proposed framework is able of producing the detected matter excess when the decoupling temperature and the inflationary energy scale (\(M_I\)) satisfy condition \(T_D \leq M_I\). Observational estimates based on gravitational-wave signals recorded by LIGO suggest that \(M_I\) is approximately \(M_I \simeq 2\times10^{16}\,\mathrm{GeV}\) \cite{oikonomou2016f,lambiase2006baryogenesis,ramos2017baryogenesis}. During the estimation of Eq. \eqref{eq:ratio_expsn_GB}, the parameter values are chosen as \(g_b \simeq 1\), \(g_*=106\), \(M_*=2\times10^{18}\,\mathrm{GeV}\), and \(T_D=2\times10^{16}\,\mathrm{GeV}\) \cite{davoudiasl2004gravitational,nozari2018baryogenesis,ramos2017baryogenesis}.

\begin{figure}[htbp]
\centering
  \includegraphics[width=0.90\linewidth]{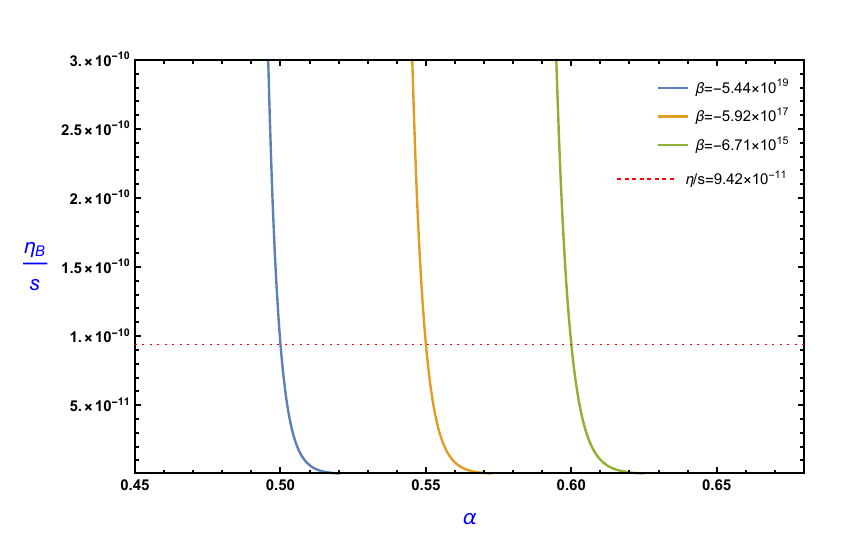}
  \caption{\justifying Plot of \(\frac{\eta_B}{s}\) versus \(\alpha\) for the \(f(R,\xi)\) model under consideration in radiation era \((\omega=\frac{1}{3})\) for three different \(\beta\) choices.}
  \label{fig:Figure2}
\end{figure}

Fig. \ref{fig:Figure2} illustrates the evolution of the generalized BnER \(\left(\frac{\eta_B}{s}\right)\) versus \(\alpha\) for three selected \(\beta\) values, namely \(\beta=-5.44 \times 10^{19}\), \(\beta=-5.92 \times 10^{17}\), and \(\beta=-6.71 \times 10^{15}\). The observationally determined baryon asymmetry is marked by the red line (dashed).  A comparison between theoretical predictions and observational bounds indicates that the parameter \(\alpha\) remains feasible roughly in the range \(0.5 \leq \alpha \leq 0.6\), demonstrating good agreement between the model predictions and cosmic observation.

\section{Generalized Gravitational Baryogenesis in \(f(R, \xi)\)}
\label{sec4}

In this section, we generalize the analysis by incorporating a more complete \(\mathcal{CP}\)-non-conserving interplay within the paradigm of the adopted \(f(R,\xi)\) model. In the standard GB case, the coupling factor is explicitly associated with \((R+\xi)\). However, in the generalized scenario, the coupling dynamics is taken to be directly related to the model \(f(R,\xi)\) itself, leading to a richer \(\mathcal{CP}\)-breaking formulation. As a result, the interaction factor governing this generalized setup takes the form \cite{nozari2018baryogenesis}:

\begin{equation}
\frac{1}{M_*^2}
\int d^4x \,\sqrt{-g}\; \partial_{\mu}f(R,\xi ) J^{\mu}\,.
\label{eq:modified_coupling_equation_GGB}
\end{equation}

Using Eq.~\eqref{eq:modified_coupling_equation_GGB}, the GGB framework yields the following general form for \(\frac{\eta_B}{s}\):

\begin{equation}
\frac{\eta_B}{s} \cong -\frac{15}{4\pi^2}\frac{g_b}{g_*}\frac{(\dot{R} f_R + \dot{\xi} f_{\xi})}{M_*^2T_D}.
\label{eq:general_baryontoentropyratio_GGB}
\end{equation}

Substituting Eqs.~\eqref{eq:ricci_scalar}, \eqref{eq:power_law_a}, \eqref{eq:f_form} and \eqref{eq:tD_for_GB} into Eq.~\eqref{eq:general_baryontoentropyratio_GGB} leads to the following expression for \(\frac{\eta_B}{s}\):

\begin{equation}
\begin{split}
\frac{\eta_B}{s}\;\Big|_{GGB} &= -\left(3^{\alpha -\frac{3}{2 \alpha }+1} \alpha  \beta  \left(r^{\alpha -1} \pi ^{-2 (1-\alpha )} (1-4 \alpha ) \beta \right)^{-\frac{3}{2 \alpha }} \left(-3^{\alpha } (4 \alpha -1) \beta  \left(3^{\frac{1}{2 \alpha }} \left(r^{\alpha -1} \pi ^{-2 (1-\alpha )} (1-4 \alpha ) \beta \right)^{\frac{1}{2 \alpha }}\right)^{-2 \alpha }\right)^{\frac{1}{1-\alpha }} \right)  \\ & \times \frac{\left(3^{-1/\alpha } \left(r^{\alpha -1} \pi ^{-2 (1-\alpha )} (1-4 \alpha ) \beta \right)^{-1/\alpha } \left(3^{\alpha } (1-4 \alpha ) \beta  \left(3^{\frac{1}{2 \alpha }} \left(r^{\alpha -1} \pi ^{-2 (1-\alpha )} (1-4 \alpha ) \beta \right)^{\frac{1}{2 \alpha }}\right)^{-2 \alpha }\right)^{\frac{1}{1-\alpha }}\right)^{\alpha -1}}{r \ \pi ^2 (\alpha -1)},
\label{eq:ratio_for_GGB}
\end{split}
\end{equation}

where,

$$r=16.96 \times 10^{37}$$

For the investigation of GB in the  generalized formalism  within the \(f(R, \xi)\) gravity, the following fixed parameters are adopted in Eq.~\eqref{eq:ratio_for_GGB}: \(g_* = 106\), \(g_b \simeq 1\), \(M_* = 2 \times 10^{18} \, \text{GeV}\), and \(T_D = 2 \times 10^{16} \, \text{GeV}\).

\begin{figure}[htbp]
\centering
  \includegraphics[width=0.90\linewidth]{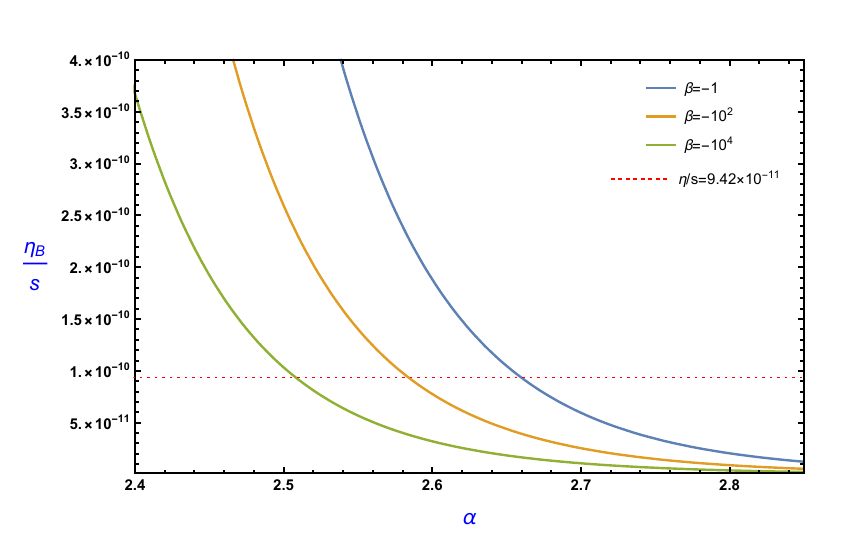}
  \caption{\justifying Plot of \(\frac{\eta_B}{s}\) versus \(\alpha\) (generalized case) for the \(f(R,\xi)\) model under consideration in radiation era \((\omega=\frac{1}{3})\) for three different \(\beta\) choices.}
  \label{fig:Figure6}
\end{figure}

Fig. \ref{fig:Figure6} shows a variation of \((\frac{\eta_B}{s})_{ggb}\) as a function of \(\alpha\) for the adopted values of \(\beta\), namely \(\beta=-1\), \(\beta=-10^{2}\) and \(\beta=-10^{4}\). The observationally determined baryon asymmetry is marked by the red line (dashed). Theoretical results compared to observational limits conclude that the viable range of \(\alpha\) lies approximately within the interval: (\(2.5, 2.65\)), indicating a close alignment between the theoretical analysis and the observational findings.

\begin{table}[htbp]
\centering
\caption{Table presenting different pairs of the parameters $(\alpha,\beta)$ in the GGB scenario, used to investigate the generation of the baryon asymmetry ratio.}
\label{tab:GGB_Table}
\begin{ruledtabular}
\begin{tabular}{ccc}
\textbf{$\alpha$} & \textbf{$\beta$} & \textbf{$\frac{\eta_B}{s}$} \\ 
\hline
 $\alpha < 1$ & Any value of $\beta$ & Negative value of the asymmetry,  \\
            &               &   implying anti-matter generation.\\
 $\alpha=1$ & Any value of $\beta$ & Indeterminate \\
 $\alpha=1.5$ & $\beta=-4 \times 10^{30}$ & $9.58414 \times 10^{-11}$ \\
 $\alpha=2$ & $\beta=-2.5 \times 10^{17}$ & $9.33407 \times 10^{-11}$ \\
 $\alpha=2.659$ & $\beta=-1$ & $9.44735 \times 10^{-11}$ \\
 $\alpha=3$ & $\beta=-1 \times 10^{-9}$ & $9.41567 \times 10^{-11}$ \\
 $\alpha=5$ & $\beta=-6.15 \times 10^{-63}$ & $9.42687 \times 10^{-11}$ \\
\end{tabular}
\end{ruledtabular}
\end{table}

Table~\ref{tab:GGB_Table} lists several viable combinations of the parameters \((\alpha,\beta)\) to examine the validity of the ratio for different combinations.

\section{Observational Viability of the Parameter Space}
\label{sec5}

The compatibility of the model parameters, obtained within the framework of $f(R,\xi)$ gravity and consistent with the generation of the observed baryon asymmetry, is tested against two well-established cosmological datasets: the Cosmic Chronometer (CC) dataset \cite{Samaddar2026Mar} and the Pantheon+SH0ES supernova compilation \cite{Camlibel2020Oct,Scolnic2022Oct}. To quantitatively assess the alignment between theoretical results and observations, a chi-square ($\chi^2$) statistical analysis is employed \cite{Sultan2025Mar}.

For the CC dataset that directly measures the Hubble parameter $ H(z)$, the expression $\chi^2$ is written as \cite{Mazumdar2026Apr}:

\begin{equation}
\chi^2_{CC} = \sum_{i=1}^{31} \frac{\left[H_{\text{th}}(z_i, \alpha, \beta) - H_{\text{obs}}(z_i)\right]^2}{\sigma_{H,i}^2},
\end{equation}

where $H_{\text{th}}(z_i, \alpha, \beta)$ represents the calculated value of $H(z)$ at $z_i$, $H_{\text{obs}}(z_i)$ is the measured magnitude, $\sigma_{H,i}$ represents the corresponding uncertainty.

For the SH0ES+Pantheon dataset, which constrains the expansion history through luminosity distance measurements, the $\chi^2$ parameter is defined as \cite{Samaddar2025Mar}:

\begin{equation}
\chi^2_{SN} = \sum_{i=1}^{1701} \frac{\left[\mu_{\text{th}}(z_i, \theta) - \mu_{\text{obs}}(z_i)\right]^2}{\sigma_{\mu,i}^2}.
\end{equation}

The distance modulus is computed theoretically from the luminosity distance as \cite{Mazumdar2025Oct}:

\begin{equation}
\mu(z) = 5 \log_{10} \left( \frac{d_L(z)}{H_0 M  pc} \right) + 25,
\end{equation}

with,

\begin{equation}
d_L(z) = (1+z)\int_0^z \frac{dz'}{H(z')}.
\end{equation}

The total chi-square is constructed as \cite{Samaddar2025Mar}:

\begin{equation}
\chi^2_{\text{total}} = \chi^2_{CC} + \chi^2_{SN},
\label{eq:total_chi_square}
\end{equation}

which is minimized to obtain the best-fit parameter values \cite{Mazumdar2025Oct}. This combined analysis enables a direct comparison between the predictions of the $f(R,\xi)$ framework and the standard $\Lambda$CDM case. The statistical deviation from $\Lambda$CDM is quantified through the resulting $\chi^2$ values and the associated confidence intervals.

This framework provides robust constraints on the model parameters and facilitates an assessment of its potential to simultaneously accommodate the late-time cosmic expansion and the observed baryon asymmetry.

The Lagrangian in Eq. \eqref{eq:f_form} is inserted into Eq. \eqref{eq:friedmann_equation1} to obtained the First Order $H(z)$ Differential Equation (DE) of the following form:

\begin{equation}
\begin{split}
& H'(z)+\frac{9^{\alpha } \alpha  \beta  (z+1) \left(H_0 (z+1)^4 H^2(z) \right)^{\alpha }-36 (z+1) H^4(z)}{36 (z+1)^2 H^3(z)} + \\ &
\frac{\sqrt{(z+1)^2 \left(81^{\alpha } \alpha ^2 \beta ^2 \left(H_0 (z+1)^4 H^2(z)\right)^{2 \alpha }+36 H^4(z) \left(-9^{\alpha } (2 \alpha -1) \beta  \left(H_0 (z+1)^4 H^2(z)\right)^{\alpha }-3 H_0 (z+1)^4\right)+1296 H^8(z)\right)}}{36 (z+1)^2 H^3(z)}
=0
\end{split}
\end{equation}

\begin{figure}[htbp]
\centering
  \includegraphics[width=0.90\linewidth]{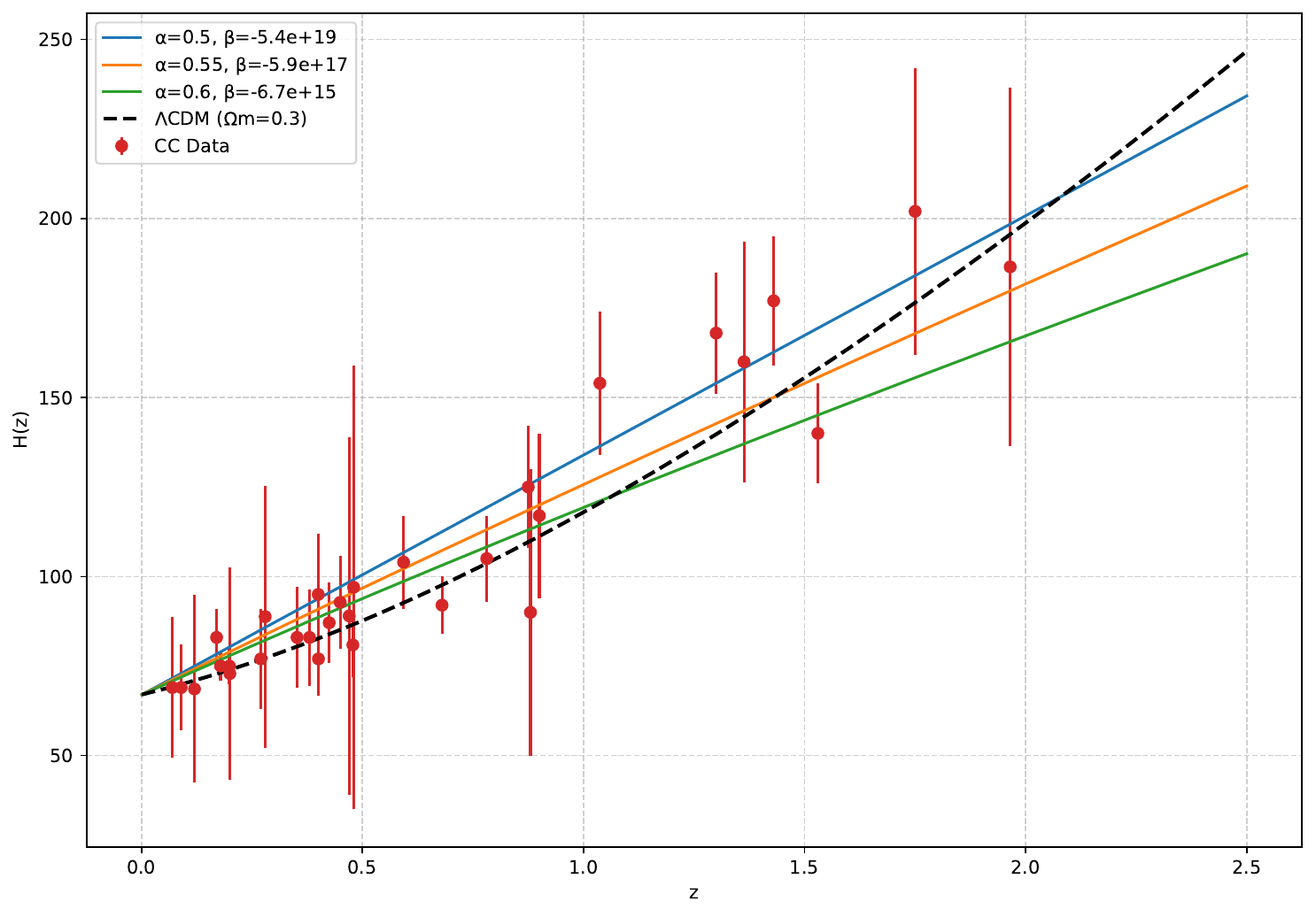}
  \caption{\justifying Evolution of $H(z)$ as a function of $z$ for different $(\alpha, \ \beta)$ values in the $f(R,\xi)$ gravity framework in comparison with $\Lambda$CDM model ($\Omega_m = 0.3$). The observational data points with error bars are taken from the 31-point CC dataset.}
  \label{fig:H(z)_vs_z}
\end{figure}

\begin{figure}[htbp]
\centering
  \includegraphics[width=0.90\linewidth]{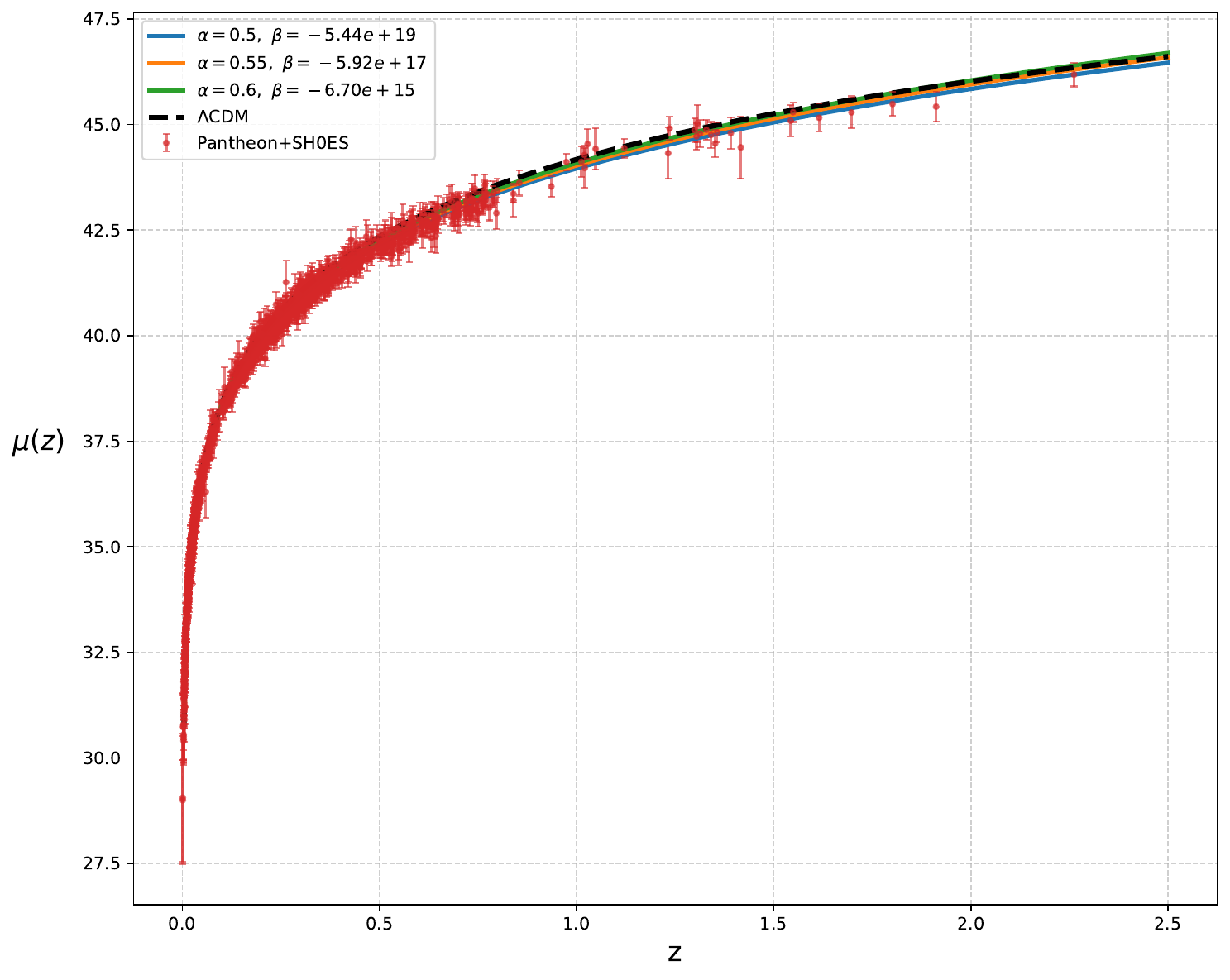}
  \caption{\justifying Variation of $\mu(z)$ as a function of $z$ for different $(\alpha, \ \beta)$ values in the $f(R,\xi)$ gravity framework in comparison with $\Lambda$CDM model ($\Omega_m = 0.3$). The observational data points with error bars are taken from the 1701-point SH0ES+Pantheon dataset.}
  \label{fig:Mu_z_vs_z}
\end{figure}

The observational validity of the adopted $f(R,\xi)$ gravity is examined by a comparison with the CC dataset and the Pantheon+SH0ES compilation. The behavior of $H(z)$ and $\mu(z)$ as functions of $z$, are presented in Figs.~\ref{fig:H(z)_vs_z} and \ref{fig:Mu_z_vs_z}, respectively.

From Fig.~\ref{fig:H(z)_vs_z}, we have plotted the variation $H(z)$ in relation to $z$ in the $f(R,\xi)$ framework. The theoretical predictions of $H(z)$ for different parameter choices $(\alpha, \beta)$ exhibit a monotonically increasing trend with redshift, compatible with the expected cosmic acceleration. The model curves corresponding to $\alpha = 0.5$, $0.55$, and $0.6$, closely follow the observational CC data points within the error bars. In particular, case $\alpha = 0.55$ shows the best agreement with the redshift range, indicating an optimal fit to the expansion rate data. Furthermore, all model trajectories remain in close proximity to the standard $\Lambda$CDM curve ($\Omega_m = 0.3$), suggesting that the parameters obtained from the gravitational baryogenesis constraint can successfully reproduce the late-time cosmic dynamics without significant deviation.

Fig.~\ref{fig:Mu_z_vs_z} illustrates the evolution of $\mu(z)$ with redshift. The theoretical curves demonstrate the expected behavior, reflecting the luminosity distance evolution in an expanding Universe. The model predictions are in close alignment with the Pantheon+SH0ES dataset throughout the redshift interval $0.07 \leq z \leq 2.5$. In contrast to the $H(z)$ vs $z$ analysis, the parameter choice $\alpha = 0.50$ provides the closest match to the observational data, while the other set of parameters also remain within the observational uncertainties. In particular, the parameter $\alpha = 0.50$ shows the closest agreement with the supernova measurements, as  supported by the minimum chi-square value. Within the parameter space restricted by gravitational baryogenesis, the $f(R,\xi)$ model is capable of describing the Pantheon+SH0ES observations more effectively than the $\Lambda$CDM model.

\begin{table}[h]
\centering
\caption{Chi-square ($\chi^2$) values for different parameter choices using the Cosmic Chronometer (CC) and Pantheon+SH0ES datasets.}
\begin{tabular}{c c c c c}
\hline
$\alpha$ & $\beta$ & $\chi^2_{CC}$ & $\chi^2_{SN}$ & $\chi^2_{\text{total}}$ \\
\hline
\hline
$\alpha = 0.50$ & $\beta= -5.44 \times 10^{19}$ & 27.085 & 1076.600 & 1096.685 \\
$\alpha = 0.55$ & $\beta= -5.92 \times 10^{17}$ & 20.313 & 1185.892 & 1206.205 \\
$\alpha = 0.60$ & $\beta= -6.70 \times 10^{15}$ & 20.937 & 1311.788 & 1332.725 \\
$\Lambda$CDM &    & 16.271  & 1905.132 & 1921.403 \\
\hline
\end{tabular}
\label{tab:chi2_combined}
\end{table}

The observational viability of the model is assessed by a statistical analysis based on the $\chi^2$ values derived for both datasets, as summarized in Table~\ref{tab:chi2_combined}. For the CC dataset, the minimized $\chi^2$ measure is obtained for the $\Lambda$CDM model, while the $f(R,\xi)$ model yields slightly larger values, with $\alpha = 0.55$ and $\alpha = 0.60$ providing a comparatively better fit. On the other hand, for the Pantheon+SH0ES dataset, the $f(R,\xi)$ model produces significantly lower chi-square values compared to $\Lambda$CDM. The minimum value is obtained for $\alpha = 0.50$, suggesting that this parameter choice is most compatible with supernova observations.

To obtain a comprehensive assessment of the performance of the model, the chi-square values of both CC and Pantheon+SH0ES datasets are combined, as shown in Table~\ref{tab:chi2_combined}. The total chi-square is defined in Eq. \eqref{eq:total_chi_square}. From the combined analysis, it is evident that the $f(R,\xi)$ model offers an improved fit to the observations compared to the $\Lambda$CDM model. In particular, the minimum total chi-square is obtained for $(\alpha,\beta) = (0.50,-5.44 \times 10^{19})$, indicating that this parameter choice is the most favored when both datasets are considered simultaneously. In general, the findings suggest that the $f(R,\xi)$ model demonstrates the ability to account for gravitational baryogenesis within the parameter space considered, while remaining compatible with the observational data.

\section{Conclusion}
\label{sec6}

The source of the measured baryon surplus in the cosmos remains an unresolved and vigorously contested issue within the paradigm of theoretical astrophysics \cite{dine2003origin}. GB represents one of the physically appealing proposals to resolve this enduring issue, as it links the generation of the baryon–antibaryon imbalance directly to the geometry of spacetime rather than to the processes of particle physics alone \cite{davoudiasl2004gravitational,lambiase2006baryogenesis}. In this work, we examine the mechanism of GB within the $f(R, \xi)$ formulation, a framework that has already demonstrated substantial potential in the broader cosmological context. Specifically, previous work has established that the $f(R, \xi)$ model encapsulates fundamental dynamics across several epochs of cosmic history, effectively replicating the observed sequence of late-time cosmological evolution \cite{Marciu2023Oct}. Additionally, it has been argued in \cite{Marciu2023Oct} that the theoretical framework $f(R,\xi)$ offers a robust theoretical setting to address a range of long-standing discrepancies in cosmology.

The emergence of the measured baryonic excess in the cosmos can be linked to \(\mathcal{CP}\)-non-conserving couplings, characterized by derivative operators such as \(\partial_{\mu}(R+\xi)\) and \(\partial_{\mu} f(R,\xi)\), which serve as a distinct dynamical driver of matter imbalance within this framework. Our investigation yields baryon-to-entropy ratios $\frac{\eta_B}{s}$ that are consistent with current observational bounds, specifically, $ \frac{\eta_B}{s} \cong 9.42 \times 10^{-11} $. This agreement indicates that the $f(R,\xi)$ framework offers a plausible mechanism to explain the observed baryon asymmetry. We further extend the analysis to a generalized scenario in which the original interaction term is modified to incorporate additional degrees of freedom, broadening the theoretical scope of the model. The corresponding results remain in accordance with observational constraints, further supporting the robustness of the proposed approach.

Major findings of this study are outlined below:

\begin{enumerate}

 \item  Key outcomes of GB analysis:

\begin{itemize}

\item The analysis of the gravity model \(f(R,\xi)\) highlights that the predicted BnER $(\frac{\eta_B}{s})$ remains compatible with current observational constraints for the parameter choices $\beta = -5.44\times 10^{19}, \ \beta = -5.92 \times 10^{17}, $ and $ \beta = -6.71 \times 10^{15}$. The evolution of $H(z)$ examined against the CC data and the standard $\Lambda$CDM framework, demonstrates that the model can successfully regenerate the observed cosmic acceleration within the parameter bounds constrained by gravitational baryogenesis. The comparison with the Pantheon+SH0ES supernova compilation further indicates that the $f(R,\xi)$ framework offers a plausible description of the observational data through the $\mu(z)$ versus $z$ analysis for different choices of $\alpha$ and $\beta$ \cite{bennett2003microwave,burles2001big}.  Thus, the \(f(R,\xi)\) formalism emerges as a reliable candidate to explain the observed matter imbalance in our cosmos.

\item For the observational range of the matter imbalance, $ 4.8 \times 10^{-11} \leq \frac{\eta_B}{s} \leq 1.03 \times 10^{-10} $ \cite{White2022Apr}, the parameters  $\alpha$ and $\beta$ become constrained to ranges within which the asymmetry can be realized. By imposing the bound $ 0.5 \leq \alpha \leq 0.6 $, the parameter $\beta$ is restricted to the interval $ -6.75 \times 10^{19} \leq \beta \leq -6.25 \times 10^{15} $.

\end{itemize}

 \item  Key outcomes of GGB analysis:

\begin{itemize} 

\item We have further extended the analysis to the generalized (GGB) scenario. Eq. ~\eqref{eq:ratio_for_GGB} represents the asymmetry ratio corresponding to the GGB case. In Fig.~\ref{fig:Figure6}, the BnER is plotted in relation to $\alpha$ for three different choices of $\beta$, namely $\beta = -1, \beta=-10^2$ and $\beta = -10^4$. The obtained values of the asymmetry ratio are shown to align closely with the observational bounds.

\item It is observed that for $\alpha \leq 1$, the model fails to generate the required asymmetry ratio for any choice of $\beta$. However, once the condition $\alpha > 1$ is satisfied, suitable combinations of the parameters  $(\alpha, \beta)$  successfully reproduce the observed matter asymmetry. Several of such viable combinations $(\alpha, \beta)$ are listed in Table~\ref{tab:GGB_Table}.

\end{itemize}

 \item  Key outcomes of the statistical analysis:

\begin{itemize} 

\item The joint assessment of the CC and Pantheon+SH0ES datasets demonstrates that the adopted gravity formulation is observationally viable for parameter space restricted by GB and compatible with the cosmic acceleration of the cosmos. The model effectively recreates the behavior of both $H(z)$ and $\mu(z)$, remaining close to the $\Lambda$CDM predictions across the redshift range considered, $0.07 \leq z \leq 2.5$. Although $\Lambda$CDM gives a slightly better fit to the CC data, the $f(R,\xi)$ model significantly outperforms it for the supernova data set. The joint analysis $\chi^2$ indicates that the parameter set $(\alpha,\beta)=(0.50,-5.44 \times 10^{19})$, yields the best overall fit. These results suggest that the model can effectively account for GB while maintaining strong agreement with observational constraints.

\end{itemize}

\end{enumerate}

The findings of this work conclude that the modifications to the gravitational dynamics gravity framework furnish a physically consistent and observationally viable setting for realizing gravitational baryogenesis within the constrained region of the model parameters. The choice \(f(R,\xi)=R^2+\beta \xi^{\alpha}\) serves as an analytically manageable step in systematically probing the physical consequences of explicit matter–geometry couplings within the baryogenesis context. The distinctive novelty of the model arises primarily from its direct coupling to the term \(\xi\), which introduces a new geometric modification to the gravitational dynamics and explores a distinct class of baryogenesis framework. The present study may be further generalized by considering alternative functional forms of $f(R,\xi)$ gravity as well as different cosmological scale-factor parameterization to explore a wider range of cosmological scenarios.

\bibliography{references}

@article{Anderson1933Mar,
	author = {Anderson, Carl D.},
	title = {{The Positive Electron}},
	journal = {Phys. Rev.},
	volume = {43},
	number = {6},
	pages = {491--494},
	year = {1933},
	month = mar,
	publisher = {American Physical Society},
	doi = {10.1103/PhysRev.43.491}
}

@article{Dirac1928Feb,
	author = {Dirac, Paul Adrien Maurice},
	title = {{The quantum theory of the electron}},
	journal = {Proc. R. Soc. London A},
	volume = {117},
	number = {778},
	pages = {610--624},
	year = {1928},
	month = feb,
	issn = {0950-1207},
	publisher = {The Royal Society},
	doi = {10.1098/rspa.1928.0023}
}

@article{Aghanim2020Sep,
	author = {Planck Collaboration},
	title = {{Planck 2018 results - VI. Cosmological parameters}},
	journal = {Astron. Astrophys.},
	volume = {641},
	pages = {A6},
	year = {2020},
	month = sep,
	issn = {0004-6361},
	publisher = {EDP Sciences},
	doi = {10.1051/0004-6361/201833910}
}

@article{dolgov1988cosmology,
	author = {Dolgov, Aleksandr D. and Zel'dovich, Iakov B. and Sazhin, Mikhail V.},
	title = {{Cosmology of the early universe}},
	journal = {Moscow Izdatel Moskovskogo Universiteta Pt},
	year = {1988},
	url = {https://ui.adsabs.harvard.edu/abs/1988MIzMU.........D/abstract}
}

@article{Peebles2003Apr,
	author = {Peebles, P. J. E. and Ratra, Bharat},
	title = {{The cosmological constant and dark energy}},
	journal = {Rev. Mod. Phys.},
	volume = {75},
	number = {2},
	pages = {559--606},
	year = {2003},
	month = apr,
	publisher = {American Physical Society},
	doi = {10.1103/RevModPhys.75.559}
}

@book{weinberg1995quantum,
	author = {Weinberg, Steven},
	title = {{The Quantum Theory of Fields}},
	year = {1995},
    publisher={Cambridge university press},
	month = jun,
	isbn = {978-052155001},
	doi = {10.1017/CBO9781139644167}
}

@book{Cheng1984,
	author = {Cheng, Ta-Pei and Li, Ling-Fong},
	title = {{Gauge Theory of Elementary Particle Physics}},
	year = {1984},
	publisher = {Oxford University Press},
	isbn = {9780198519614},
	doi = {20.500.12657/59106}
}

@article{bennett2003microwave,
	author = {Bennett, C. L. and Bay, M. and Halpern, M. and Hinshaw, G. and Jackson, C. and Jarosik, N. and Kogut, A. and Limon, M. and Meyer, S. S. and Page, L. and Spergel, D. N. and Tucker, G. S. and Wilkinson, D. T. and Wollack, E. and Wright, E. L.},
	title = {{The Microwave Anisotropy Probe{$\ast$} Mission}},
	journal = {Astrophys. J.},
	volume = {583},
	number = {1},
	pages = {1},
	year = {2003},
	month = jan,
	issn = {0004-637X},
	publisher = {IOP Publishing},
	doi = {10.1086/345346}
}

@article{Zeldovich1968Mar,
	author = {Zel{'}dovich, Yakov Borisovich},
	title = {{The cosmological constant and the theory of elementary particles}},
	journal = {Phys.-Usp.},
	volume = {11},
	number = {3},
	pages = {381--393},
	year = {1968},
	month = mar,
	issn = {1063-7869},
	url = {https://ufn.ru/en/articles/1968/3/m}
}

@article{davoudiasl2004gravitational,
	author = {Davoudiasl, Hooman and Kitano, Ryuichiro and Kribs, Graham D. and Murayama, Hitoshi and Steinhardt, Paul J.},
	title = {{Gravitational Baryogenesis}},
	journal = {Phys. Rev. Lett.},
	volume = {93},
	number = {20},
	pages = {201301},
	year = {2004},
	month = nov,
	publisher = {American Physical Society},
	doi = {10.1103/PhysRevLett.93.201301}
}

@article{burles2001big,
	author = {Burles, Scott and Nollett, Kenneth M. and Turner, Michael S.},
	title = {{What is the big-bang-nucleosynthesis prediction for the baryon density and how reliable is it?}},
	journal = {Phys. Rev. D},
	volume = {63},
	number = {6},
	pages = {063512},
	year = {2001},
	month = feb,
	publisher = {American Physical Society},
	doi = {10.1103/PhysRevD.63.063512}
}

@article{steigman1976observational,
	author = {Steigman, Gary},
	title = {{Observational Tests of Antimatter Cosmologies}},
	journal = {Annual Review of Astronomy and Astrophysics},
	number = {Volume 14, 1976},
	pages = {339--372},
	year = {1976},
	month = sep,
	publisher = {Annual Reviews},
	doi = {10.1146/annurev.aa.14.090176.002011}
}

@article{dolgov2001matter,
	author = {Dolgov, A.},
	title = {{Matter-antimatter domains in the universe}},
	journal = {Nucl. Phys. B Proc. Suppl.},
	volume = {95},
	number = {1},
	pages = {42--46},
	year = {2001},
	month = apr,
	issn = {0920-5632},
	publisher = {North-Holland},
	doi = {10.1016/S0920-5632(01)01051-9}
}

@article{cohen1998matter,
	author = {Cohen, A. G. and De R{\ifmmode\acute{u}\else\'{u}\fi}jula, A. and Glashow, S. L.},
	title = {{A Matter-Antimatter Universe?}},
	journal = {Astrophys. J.},
	volume = {495},
	number = {2},
	pages = {539},
	year = {1998},
	month = mar,
	issn = {0004-637X},
	publisher = {IOP Publishing},
	doi = {10.1086/305328}
}

@article{dine2003origin,
	author = {Dine, Michael and Kusenko, Alexander},
	title = {{Origin of the matter-antimatter asymmetry}},
	journal = {Rev. Mod. Phys.},
	volume = {76},
	number = {1},
	pages = {1--30},
	year = {2003},
	month = dec,
	publisher = {American Physical Society},
	doi = {10.1103/RevModPhys.76.1}
}

@book{Kolb2018Mar,
	author = {Kolb, Edward},
	title = {{The Early Universe}},
	journal = {Taylor {\&} Francis},
	year = {2018},
	month = mar,
	isbn = {978-0-42949286-0},
	publisher = {Taylor {\&} Francis},
	address = {Andover, England, UK},
	doi = {10.1201/9780429492860}
}

@article{sakharov1998violation,
	author = {Sakharov, Andrei D.},
	title = {{Violation of CP invariance, C asymmetry, and baryon asymmetry of the universe}},
	journal = {Sov. Phys. Usp.},
	volume = {34},
	number = {5},
	pages = {392},
	year = {1991},
	month = may,
	issn = {0038-5670},
	publisher = {IOP Publishing},
	doi = {10.1070/PU1991v034n05ABEH002497}
}

@article{brandenberger2003spontaneous,
	author = {Brandenberger, Robert H. and Yamaguchi, Masahide},
	title = {{Spontaneous baryogenesis in warm inflation}},
	journal = {Phys. Rev. D},
	volume = {68},
	number = {2},
	pages = {023505},
	year = {2003},
	month = jul,
	publisher = {American Physical Society},
	doi = {10.1103/PhysRevD.68.023505}
}

@article{takahashi2004spontaneous,
	author = {Takahashi, Fuminobu and Yamaguchi, Masahide},
	title = {{Spontaneous baryogenesis in flat directions}},
	journal = {Phys. Rev. D},
	volume = {69},
	number = {8},
	pages = {083506},
	year = {2004},
	month = apr,
	publisher = {American Physical Society},
	doi = {10.1103/PhysRevD.69.083506}
}

@article{trodden1999electroweak,
	author = {Trodden, Mark},
	title = {{Electroweak baryogenesis}},
	journal = {Rev. Mod. Phys.},
	volume = {71},
	number = {5},
	pages = {1463--1500},
	year = {1999},
	month = oct,
	publisher = {American Physical Society},
	doi = {10.1103/RevModPhys.71.1463}
}

@article{kolb1996grand,
	author = {Kolb, Edward W. and Linde, Andrei and Riotto, Antonio},
	title = {{Grand-Unified-Theory Baryogenesis after Preheating}},
	journal = {Phys. Rev. Lett.},
	volume = {77},
	number = {21},
	pages = {4290--4293},
	year = {1996},
	month = nov,
	publisher = {American Physical Society},
	doi = {10.1103/PhysRevLett.77.4290}
}

@article{stewart1996affleck,
	author = {Stewart, E. D. and Kawasaki, M. and Yanagida, T.},
	title = {{Affleck-Dine baryogenesis after thermal inflation}},
	journal = {Phys. Rev. D},
	volume = {54},
	number = {10},
	pages = {6032--6039},
	year = {1996},
	month = nov,
	publisher = {American Physical Society},
	doi = {10.1103/PhysRevD.54.6032}
}

@article{ambrosone2022towards,
	author = {Ambrosone, Antonio and Calabrese, Roberta and Fiorillo, Damiano F. G. and Miele, Gennaro and Morisi, Stefano},
	title = {{Towards baryogenesis via absorption from primordial black holes}},
	journal = {Phys. Rev. D},
	volume = {105},
	number = {4},
	pages = {045001},
	year = {2022},
	month = feb,
	publisher = {American Physical Society},
	doi = {10.1103/PhysRevD.105.045001}
}

@article{Luty1992Jan,
	author = {Luty, Markus A.},
	title = {{Baryogenesis via leptogenesis}},
	journal = {Phys. Rev. D},
	volume = {45},
	number = {2},
	pages = {455--465},
	year = {1992},
	month = jan,
	publisher = {American Physical Society},
	doi = {10.1103/PhysRevD.45.455}
}

@article{Barbieri2000May,
	author = {Barbieri, Riccardo and Creminelli, Paolo and Strumia, Alessandro and Tetradis, Nikolaos},
	title = {{Baryogenesis through leptogenesis}},
	journal = {Nucl. Phys. B},
	volume = {575},
	number = {1},
	pages = {61--77},
	year = {2000},
	month = may,
	issn = {0550-3213},
	publisher = {North-Holland},
	doi = {10.1016/S0550-3213(00)00011-0}
}

@article{Marciu2023Oct,
	author = {Marciu, Mihai and Ioan, Dana Maria},
	title = {{Physical aspects of ${ \boldsymbol{f}(\boldsymbol R,\boldsymbol G_{\boldsymbol\mu \boldsymbol\nu}\boldsymbol T^{\boldsymbol\mu \boldsymbol\nu})} $ modified gravity theories}},
	journal = {Chin. Phys. C},
	volume = {47},
	number = {10},
	pages = {105103},
	year = {2023},
	month = oct,
	issn = {1674-1137},
	publisher = {Chinese Physical Society and the Institute of High Energy Physics of the Chinese Academy of Sciences and the Institute of Modern Physics of the Chinese Academy of Sciences and IOP Publishing Ltd},
	doi = {10.1088/1674-1137/ace81e}
}

@article{lambiase2006baryogenesis,
	author = {Lambiase, G. and Scarpetta, G.},
	title = {{Baryogenesis in $f(R)$ theories of gravity}},
	journal = {Phys. Rev. D},
	volume = {74},
	number = {8},
	pages = {087504},
	year = {2006},
	month = oct,
	publisher = {American Physical Society},
	doi = {10.1103/PhysRevD.74.087504}
}

@article{Oikonomou2016Dec,
	author = {Oikonomou, V. K. and Saridakis, Emmanuel N.},
	title = {{$f(T)$ gravitational baryogenesis}},
	journal = {Phys. Rev. D},
	volume = {94},
	number = {12},
	pages = {124005},
	year = {2016},
	month = dec,
	publisher = {American Physical Society},
	doi = {10.1103/PhysRevD.94.124005}
}

@article{Odintsov2016Sep,
	author = {Odintsov, S. D. and Oikonomou, V. K.},
	title = {{Gauss{\textendash}Bonnet gravitational baryogenesis}},
	journal = {Phys. Lett. B},
	volume = {760},
	pages = {259--262},
	year = {2016},
	month = sep,
	issn = {0370-2693},
	publisher = {North-Holland},
	doi = {10.1016/j.physletb.2016.06.074}
}

@article{Oikonomou2016Mar,
	author = {Oikonomou, V. K.},
	title = {{Constraints on singular evolution from gravitational baryogenesis}},
	journal = {Int. J. Geom. Methods Mod. Phys.},
	volume = {13},
	number = {03},
	year = {2016},
	month = mar,
	publisher = {World Scientific Publishing Company},
	doi = {10.1142/S021988781650033X}
}

@article{sahoo2020gravitational,
	author = {Sahoo, P. K. and Bhattacharjee, Snehasish},
	title = {{Gravitational Baryogenesis in Non-Minimal Coupled f(R,T) Gravity}},
	journal = {Int. J. Theor. Phys.},
	volume = {59},
	number = {5},
	pages = {1451--1459},
	year = {2020},
	month = may,
	issn = {1572-9575},
	publisher = {Springer US},
	doi = {10.1007/s10773-020-04414-3}
}

@article{malakar2025f,
	author = {Malakar, Kalyan and Mazumdar, Rajdeep and Gohain, Mrinnoy M. and Bhuyan, Kalyan},
	title = {{f(R) Gravity reconstruction using Barrow holographic dark energy model}},
	journal = {Journal of Subatomic Particles and Cosmology},
	volume = {3},
	pages = {100068},
	year = {2025},
	month = jun,
	issn = {3050-4805},
	publisher = {Elsevier},
	doi = {10.1016/j.jspc.2025.100068}
}

@article{nozari2018baryogenesis,
	author = {Nozari, Kourosh and Rajabi, Fateme},
	title = {{Baryogenesis in f(R, T ) Gravity{$\ast$}}},
	journal = {Commun. Theor. Phys.},
	volume = {70},
	number = {4},
	pages = {451},
	year = {2018},
	month = oct,
	issn = {0253-6102},
	publisher = {Chinese Physical Society and IOP Publishing Ltd},
	doi = {10.1088/0253-6102/70/4/451}
}

@article{baffou2019f,
	author = {Baffou, E. H. and Houndjo, M. J. S. and Kanfon, D. A. and Salako, I. G.},
	title = {{f(R, T) models applied to baryogenesis}},
	journal = {Eur. Phys. J. C},
	volume = {79},
	number = {2},
	pages = {1--6},
	year = {2019},
	month = feb,
	issn = {1434-6052},
	publisher = {Springer Berlin Heidelberg},
	doi = {10.1140/epjc/s10052-019-6559-0}
}

@article{Atazadeh2018Jun,
	author = {Atazadeh, K.},
	title = {{Gravitational baryogenesis in DGP brane cosmology}},
	journal = {Eur. Phys. J. C},
	volume = {78},
	number = {6},
	pages = {455},
	year = {2018},
	month = jun,
	issn = {1434-6052},
	publisher = {Springer Berlin Heidelberg},
	doi = {10.1140/epjc/s10052-018-5952-4}
}

@article{ramos2017baryogenesis,
	author = {Ramos, M. P. L. P. and P{\ifmmode\acute{a}\else\'{a}\fi}ramos, J.},
	title = {{Baryogenesis in nonminimally coupled f(R) theories}},
	journal = {Phys. Rev. D},
	volume = {96},
	number = {10},
	pages = {104024},
	year = {2017},
	month = nov,
	publisher = {American Physical Society},
	doi = {10.1103/PhysRevD.96.104024}
}

@article{oikonomou2016f,
	author = {Oikonomou, V. K. and Saridakis, Emmanuel N.},
	title = {{f(T) gravitational baryogenesis}},
	journal = {Phys. Rev. D},
	volume = {94},
	number = {12},
	year = {2016},
	month = jul,
	issn = {2470-0010},
	publisher = {American Physical Society},
	doi = {10.1103/PhysRevD.94.124005}
}

@article{jaybhaye2023baryogenesis,
	author = {Jaybhaye, Lakhan V. and Bhattacharjee, Snehasish and Sahoo, P. K.},
	title = {{Baryogenesis in f(R,Lm) gravity}},
	journal = {Phys. Dark Universe},
	volume = {40},
	pages = {101223},
	year = {2023},
	month = may,
	issn = {2212-6864},
	publisher = {Elsevier B.V.},
	doi = {10.1016/j.dark.2023.101223}
}

@article{Samaddar2025Mar,
	author = {Samaddar, Amit and Singh, S. Surendra},
	title = {{A novel approach to baryogenesis in f(Q,Lm) gravity and its cosmological implications}},
	journal = {Nucl. Phys. B},
	volume = {1012},
	pages = {116834},
	year = {2025},
	month = mar,
	issn = {0550-3213},
	publisher = {North-Holland},
	doi = {10.1016/j.nuclphysb.2025.116834}
}

@article{Mishra2024Apr,
	author = {Mishra, Sai Swagat and Bhat, Aaqid and Sahoo, P. K.},
	title = {{Probing baryogenesis in f(Q) gravity}},
	journal = {Europhys. Lett.},
	volume = {146},
	number = {2},
	pages = {29001},
	year = {2024},
	month = apr,
	issn = {0295-5075},
	publisher = {EDP Sciences, IOP Publishing and Societ{\ifmmode\grave{a}\else\`{a}\fi} Italiana di Fisica},
	doi = {10.1209/0295-5075/ad329b}
}

@article{SwagatMishra2026Jan,
	author = {Swagat Mishra, Sai and Kavya, N. S. and Sahoo, P. K.},
	title = {{Effects of DESI and GW observations on f(T) gravitational baryogenesis}},
	journal = {Phys. Lett. B},
	volume = {872},
	pages = {140036},
	year = {2026},
	month = jan,
	issn = {0370-2693},
	publisher = {North-Holland},
	doi = {10.1016/j.physletb.2025.140036}
}

@article{Cruz2026Feb,
	author = {Cruz, Daniel F. P. and Pereira, David S. and Lobo, Francisco S. N.},
	title = {{Gravitational baryogenesis in f(T, Lm) gravity}},
	journal = {Nucl. Phys. B},
	volume = {1023},
	pages = {117304},
	year = {2026},
	month = feb,
	issn = {0550-3213},
	publisher = {North-Holland},
	doi = {10.1016/j.nuclphysb.2026.117304}
}

@article{Sultan2025Mar,
	author = {Sultan, Abdul Malik and Mushtaq, Alishba and Chou, Dean and Rehman, Hamood Ur and Ashraf, Hameed and Awan, Aziz Ullah},
	title = {{Observational analysis of gravitational baryogenesis constraints in Einstein-{\AE}ther gravity}},
	journal = {J. High Energy Astrophys.},
	volume = {45},
	pages = {135--145},
	year = {2025},
	month = mar,
	issn = {2214-4048},
	publisher = {Elsevier},
	doi = {10.1016/j.jheap.2024.11.018}
}

@article{Sultan2025May,
	author = {Sultan, Abdul Malik and Mushtaq, Alishba and Jawad, Abdul and Shaymatov, Sanjar and Saleem, Muhammad Shoaib},
	title = {{Gravitational baryogenesis analysis of observationally favored f(T,{$\phi$}) models}},
	journal = {Phys. Dark Universe},
	volume = {48},
	pages = {101936},
	year = {2025},
	month = may,
	issn = {2212-6864},
	publisher = {Elsevier},
	doi = {10.1016/j.dark.2025.101936}
}

@article{Mazumdar2026Apr,
	author = {Mazumdar, Rajdeep and Malakar, Kalyan and Bhuyan, Kalyan},
	title = {{Fractional holographic dark energy driven reconstruction of f(Q) gravity and its cosmological implications}},
	journal = {Classical Quantum Gravity},
	volume = {43},
	number = {8},
	pages = {085004},
	year = {2026},
	month = apr,
	issn = {0264-9381},
	publisher = {IOP Publishing},
	doi = {10.1088/1361-6382/ae5cf4}
}

@article{Mazumdar2025Oct,
	author = {Mazumdar, Rajdeep and Malakar, Kalyan and Gohain, Mrinnoy M. and Bhuyan, Kalyan},
	title = {{New agegraphic dark energy driven reconstruction of f(Q) gravity and its cosmological implications}},
	journal = {Classical Quantum Gravity},
	volume = {42},
	number = {20},
	pages = {205010},
	year = {2025},
	month = oct,
	issn = {0264-9381},
	publisher = {IOP Publishing},
	doi = {10.1088/1361-6382/ae0f34}
}

@article{Malakar2026Jun,
	author = {Malakar, Kalyan and Mazumdar, Rajdeep and Gohain, Mrinnoy M. and Bhuyan, Kalyan},
	title = {{Viability constraints on baryogenesis in f(R, Lm, T) gravity}},
	journal = {Phys. Dark Universe},
	volume = {52},
	pages = {102288},
	year = {2026},
	month = jun,
	issn = {2212-6864},
	publisher = {Elsevier},
	doi = {10.1016/j.dark.2026.102288}
}

@book{Mukhanov2005Nov,
	author = {Mukhanov, Viatcheslav},
	title = {{Physical Foundations of Cosmology}},
	journal = {Cambridge Core},
	year = {2005},
	month = nov,
	isbn = {978-0-52156398-7},
	publisher = {Cambridge University Press},
	address = {Cambridge, England, UK},
	doi = {10.1017/CBO9780511790553}
}

@article{rasanen2015new,
	author = {R{\ifmmode\ddot{a}\else\"{a}\fi}s{\ifmmode\ddot{a}\else\"{a}\fi}nen, Syksy and Bolejko, Krzysztof and Finoguenov, Alexis},
	title = {{New Test of the Friedmann-Lema{\ifmmode\hat{\imath}\else\^{\i}\fi}tre-Robertson-Walker Metric Using the Distance Sum Rule}},
	journal = {Phys. Rev. Lett.},
	volume = {115},
	number = {10},
	pages = {101301},
	year = {2015},
	month = sep,
	publisher = {American Physical Society},
	doi = {10.1103/PhysRevLett.115.101301}
}

@book{piattella2018lecture,
	author = {Piattella, Oliver Fabio},
	title = {{Lecture Notes in Cosmology}},
	journal = {Lecture Notes in Cosmology},
	year = {2018},
        Publisher={Springer International Publishing},
	month = sep,
	issn = {2198-7882},
	isbn = {978-3-319-95569-8},
	doi = {10.1007/978-3-319-95570-4}
}

@article{sugamoto1995baryon,
	author = {Sugamoto, Akio},
	title = {{Baryon Asymmetry: Evidence of CP Violation and Phase Transition in the Early Universe ?}},
	journal = {arXiv},
	year = {1995},
	month = may,
	eprint = {hep-ph/9505342},
	doi = {10.48550/arXiv.hep-ph/9505342}
}

@article{balaji2005dynamical,
	author = {Balaji, K. R. S. and Biswas, Tirthabir and Brandenberger, Robert H. and London, David},
	title = {{Dynamical CP violation in the early universe and leptogenesis}},
	journal = {Phys. Rev. D},
	volume = {72},
	number = {5},
	pages = {056005},
	year = {2005},
	month = sep,
	publisher = {American Physical Society},
	doi = {10.1103/PhysRevD.72.056005}
}

@article{huber2023baryogenesis,
	author = {Huber, S. J. and Mimasu, K. and No, J. M.},
	title = {{Baryogenesis from transitional CP violation in the early Universe}},
	journal = {Phys. Rev. D},
	volume = {107},
	number = {7},
	pages = {075042},
	year = {2023},
	month = apr,
	publisher = {American Physical Society},
	doi = {10.1103/PhysRevD.107.075042}
}

@article{l2017model,
	author = {L'Huillier, Benjamin and Shafieloo, Arman},
	title = {{Model-independent test of the FLRW metric, the flatness of the Universe, and non-local estimation of H0rd}},
	journal = {J. Cosmol. Astropart. Phys.},
	volume = {2017},
	number = {1},
	pages = {015},
	year = {2017},
	month = jan,
	publisher = {Institute of Physics},
	doi = {10.1088/1475-7516/2017/01/015}
}

@article{foidl2024lambda,
	author = {Foidl, Horst and Rindler-Daller, Tanja},
	title = {{A $\Lambda$CDM Extension Explaining the Hubble Tension and the Spatial Curvature $\Omega_{k,0} = -0.012 \pm 0.010$ Measured by the Final PR4 of the Planck Mission}},
	journal = {ResearchGate},
	year = {2024},
	month = dec,
	doi = {10.48550/arXiv.2412.04126}
}

@article{jimenez2019measuring,
	author = {Jimenez, Raul and Maartens, Roy and Rida Khalifeh, Ali and Caldwell, Robert R. and Heavens, Alan F. and Verde, Licia},
	title = {{Measuring the homogeneity of the universe using polarization drift}},
	journal = {J. Cosmol. Astropart. Phys.},
	volume = {2019},
	number = {5},
	pages = {048},
	year = {2019},
	month = may,
	issn = {1475-7516},
	eprint = {arXiv:1902.11298},
	doi = {10.1088/1475-7516/2019/05/048}
}

@book{White2022Apr,
	author = {White, Graham},
	title = {{Electroweak Baryogenesis (Second Edition) -- An introduction}},
	year = {2022},
	month = apr,
	isbn = {978-0-7503-3571-3},
	publisher = {IOP Publishing},
	address = {Bristol, England, UK},
	doi = {10.1088/978-0-7503-3571-3}
}

@book{Wald1984Jan,
	author = {Wald, Robert M.},
	title = {{General Relativity}},
	year = {1984},
	month = jan,
    isbn = {9780226870335},
	publisher = {The University of Chicago Press},
	doi = {10.7208/chicago/9780226870373.001.0001}
}

@article{Camlibel2020Oct,
	author = {{\ifmmode\mbox{\c{C}}\else\c{C}\fi}aml{\ifmmode\imath\else\i\fi}bel, A. Kaz{\ifmmode\imath\else\i\fi}m and Semiz, {\ifmmode\dot{I}\else\.{I}\fi}brahim and xn--Feyizolu-rwc, M. Akif},
	title = {{Pantheon update on a model-independent analysis of cosmological supernova data}},
	journal = {Classical Quantum Gravity},
	volume = {37},
	number = {23},
	pages = {235001},
	year = {2020},
	month = oct,
	issn = {0264-9381},
	publisher = {IOP Publishing},
	doi = {10.1088/1361-6382/abba48}
}

@article{Scolnic2022Oct,
	author = {Scolnic et al.,, Dan},
	title = {{The Pantheon+ Analysis: The Full Data Set and Light-curve Release}},
	journal = {Astrophys. J.},
	volume = {938},
	number = {2},
	pages = {113},
	year = {2022},
	month = oct,
	issn = {0004-637X},
	publisher = {The American Astronomical Society},
	doi = {10.3847/1538-4357/ac8b7a}
}

@article{Samaddar2026Mar,
	author = {Samaddar, Amit and Singh, S. Surendra},
	title = {{Late-time cosmic dynamics in f(R, Lm) gravity with recent observations}},
	journal = {J. High Energy Astrophys.},
	volume = {51},
	pages = {100558},
	year = {2026},
	month = mar,
	issn = {2214-4048},
	publisher = {Elsevier},
	doi = {10.1016/j.jheap.2026.100558}
}

@article{Riotto1999Dec,
	author = {Riotto, Antonio and Trodden, Mark},
	title = {{Recent progress in baryogenesis}},
	journal = {Annu. Rev. Nucl. Part. Sci.},
	number = {Volume 49, 1999},
	pages = {35--75},
	year = {1999},
	month = dec,
	publisher = {Annual Reviews},
	doi = {10.1146/annurev.nucl.49.1.35}
}

\end{document}